\documentclass[fleqn,twoside]{article}
\usepackage{espcrc2}

\usepackage{comment}

\usepackage{graphicx}
\usepackage[figuresright]{rotating}

\newcommand{\AmS}{{\protect\the\textfont2
  A\kern-.1667em\lower.5ex\hbox{M}\kern-.125emS}}

\newcommand{\lsim}
{\mathrel{\raisebox{-.3em}{$\stackrel{\displaystyle <}{\sim}$}}}
\newcommand{\gsim}
{\mathrel{\raisebox{-.3em}{$\stackrel{\displaystyle >}{\sim}$}}}
\def\asymp#1%
{\mathrel{\raisebox{-.4em}{$\widetilde{\scriptstyle #1}$}}}

\def\Nequal#1%
{\mathrel{\raisebox{-.5em}{$\stackrel{=}{\scriptstyle\rm#1}$}}}
\newcommand{\dsl}[1]{\not \hspace{-0.7mm}#1}
\def\dsl{\mathpalette\make@slash}
\def\make@slash#1#2{\setbox\z@\hbox{$#1#2$}%
  \hbox to 0pt{\hss$#1/$\hss\kern-\wd0}\box0}

\def\beq{\begin{equation}}
\def\eeq{\end{equation}}
\def\bit{\begin{itemize}}
\def\eit{\end{itemize}}
\def\beqar{\begin{eqnarray}}
\def\eeqar{\end{eqnarray}}
\def\barr#1{\begin{array}{#1}}
\def\earr{\end{array}}
\def\bfi{\begin{figure}}
\def\efi{\end{figure}}
\def\btab{\begin{table}}
\def\etab{\end{table}}
\def\bce{\begin{center}}
\def\ece{\end{center}}

\def\text{\textstyle}

\def\ga{\gamma}

\def\refeq#1{\mbox{(\ref{#1})}}

\def\reffi#1{\mbox{Figure~\ref{#1}}}

\def\citere#1{\mbox{Ref.~\cite{#1}}}
\def\citeres#1{\mbox{Refs.~\cite{#1}}}

\newcommand{\TeV}{\unskip\,\mathrm{TeV}}
\newcommand{\GeV}{\unskip\,\mathrm{GeV}}

\newcommand{\fb}{\unskip\,\mathrm{fb}}

\newcommand{\rT}{{\mathrm{T}}}

\def\mathswitchr#1{\relax\ifmmode{\mathrm{#1}}\else$\mathrm{#1}$\fi}

\newcommand{\PW}{\mathswitchr W}

\newcommand{\PZ}{\mathswitchr Z}

\newcommand{\Pg}{\mathswitchr g}

\newcommand{\PH}{\mathswitchr H}

\newcommand{\Pb}{\mathswitchr b}

\newcommand{\Pp}{\mathswitchr p}
\newcommand{\Pj}{\mathswitchr j}
\newcommand{\Pt}{\mathswitchr t}
\newcommand{\Ptbar}{\mathswitchr{\bar t}}

\newcommand{\jet}{\mathswitchr {j}}

\def\mathswitch#1{\relax\ifmmode#1\else$#1$\fi}

\newcommand{\MH}{\mathswitch {M_\PH}}

\newcommand{\Mt}{\mathswitch {m_\Pt}}

\def\solid{\raise.9mm\hbox{\protect\rule{1.1cm}{.2mm}}}
\def\dash{\raise.9mm\hbox{\protect\rule{2mm}{.2mm}}\hspace*{1mm}}

\def\ie{i.e.\ }

\title{Production of $\Pt\bar\Pt\Pb\bar\Pb$ 
at the LHC at NLO QCD
\thanks{This work is supported in part by the European
Community's Marie-Curie Research Training Network under contract
MRTN-CT-2006-035505 ``Tools and Precision Calculations for Physics
Discoveries at Colliders'' and the Japan Society for the Promotion of
Science.}}%

\author{A.\ Bredenstein\address[KEK]{High Energy Accelerator Research
                Organization (KEK),
            Tsukuba, Ibaraki 305-0801, Japan},
         A.\ Denner\address[PSI]{Paul Scherrer Institut, 
                                 W\"urenlingen und Villigen,
                                 CH-5232 Villigen PSI, Switzerland},
        S.\ Dittmaier\address[FR]{Albert-Ludwigs-Universit\"at Freiburg, 
               Physikalisches Institut, 
D-79104 Freiburg, Germany}
        and
        S.\ Pozzorini\address[CERN]{Physics Department, Theory Group, CERN,
CH-1211 Geneva 23, Switzerland}}
       
\begin{document}

\begin{abstract}
We summarise predictions for $\Pt\bar\Pt\Pb\bar\Pb$ production at the
LHC in next-to-leading order QCD.  
The precise description of this background process is a prerequisite
to observe associated $\Pt\bar\Pt\PH$ production in the
$\PH\to\Pb\bar\Pb$ decay channel.
The one-loop amplitudes are computed using Feynman diagrams and
numerical tensor reduction.  This approach provides very high
numerical stability and CPU efficiency.
We find that the scale choice adopted in ATLAS simulations underestimates the
$\Pt\bar\Pt\Pb\bar\Pb$ background by a factor two and introduce a new
dynamical scale that stabilises the perturbative predictions.
In the regime of highly boosted Higgs bosons, which offers
better perspectives to observe the $\Pt\bar\Pt\PH(\PH\to\Pb\bar\Pb)$
signal, the corrections induce significant distortions in the kinematic distributions.
\end{abstract}

\maketitle

\section{Introduction}

The discovery of the Higgs boson and the measurement of its interactions
with massive quarks and vector bosons represent a central goal of the 
Large Hadron Collider (LHC).
For a light Higgs boson,  $\MH\lsim 130 \GeV$, associated 
$\Pt\bar\Pt\PH$ production provides the opportunity to observe the
Higgs boson in the $\PH\to\Pb\bar\Pb$ decay channel and to
measure the top-quark Yukawa coupling. 
However, the extraction of the $\Pt\bar\Pt\PH(\PH\to\Pb\bar\Pb)$ signal 
from its large QCD backgrounds,
$\Pp\Pp\to\Pt\bar\Pt\Pb\bar\Pb$ and $\Pt\bar\Pt\Pj\Pj$,
represents a serious challenge.
The selection strategies elaborated by ATLAS and
CMS~\cite{Aad:2009wy,Ball:2007zza} 
anticipate a statistical significance 
around $2\sigma$ 
and a signal-to-background ratio as low as 1/10.  This calls for
better than 10\% precision in the background description, a very
demanding requirement both from the experimental and theoretical point
of view.
Very recently, a novel selection strategy
based on highly boosted Higgs bosons
has opened new
and very promising perspectives~\cite{Plehn:2009rk}.
This approach 
might 
increase the signal-to-background ratio beyond $1/3$.  Moreover, three
$\mathrm{b}$-taggings would be sufficient to strongly suppress the
$\mathrm{t\bar t jj}$ contamination so that  the background would
be completely dominated by $\mathrm{t\bar t b\bar b}$ production.

The calculation of the next-to-leading-order (NLO) QCD corrections to
the irreducible $\Pt\bar\Pt\Pb\bar\Pb$ background, first presented
in~\citeres{Bredenstein:2008zb,Bredenstein:2009aj,Bredenstein:2010rs}
and subsequently confirmed in~\citere{Bevilacqua:2009zn}, constitutes
another important step towards the observability of
$\Pt\bar\Pt\PH(\PH\to\Pb\bar\Pb)$ at the LHC.  These NLO predictions
are mandatory in order to reduce the huge scale uncertainty of the
lowest-order (LO) $\Pt\bar\Pt\Pb\bar\Pb$ cross section, which can vary
up to a factor four if the QCD scales are identified with different
kinematic parameters~\cite{Kersevan:2002vu}.
Motivated by results for the signal process $\Pp\Pp\to\Pt\bar\Pt\PH$
\cite{Beenakker:2001rj}, where a moderate $K$ factor ($K\simeq 1.2$) had
been found \cite{Beenakker:2001rj}, experimental groups adopted the
scale $\mu_\mathrm{R,F}=\Mt+m_{\Pb\bar\Pb}/2$ for
the LO simulation of the $\Pt\bar\Pt\Pb\bar\Pb$
background~\cite{Aad:2009wy}.
However, at this scale the NLO corrections to
$\Pp\Pp\to\Pt\bar\Pt\Pb\bar\Pb$ turn out to be large $(K\simeq
1.8)$~\cite{Bredenstein:2009aj,Bevilacqua:2009zn}.  

The calculation of the NLO corrections to
$\Pp\Pp\to\Pt\bar\Pt\Pb\bar\Pb$ constitutes also an important
technical benchmark.  The description of many-particle processes at
NLO plays a central role for the LHC physics programme, and the
technical challenges raised by these calculations have triggered an
impressive amount of conceptual and technical developments.
Within the last year, this progress has lead to  
the first NLO results for six-particle processes 
at the LHC, namely for 
\mbox{$\Pp\Pp\to\Pt\bar\Pt\Pb\bar\Pb$}~\cite{Bredenstein:2009aj,Bevilacqua:2009zn},
$\Pp\Pp\to\Pt\bar\Pt\Pj\Pj$ \cite{Bevilacqua:2010ve},
the leading-~\cite{KeithEllis:2009bu} and the full-colour
contributions~\cite{Berger:2009ep} to $\Pp\Pp\to\PW \Pj\Pj\Pj$, for 
$\Pp\Pp\to\PZ/\ga \Pj\Pj\Pj$  \cite{Berger:2010vm}
and for the $q\bar q$ contribution to 
$\Pp\Pp\to\Pb\bar\Pb\Pb\bar\Pb$~\cite{Binoth:2009rv}.

To compute the virtual corrections to $\Pt\bar\Pt\Pb\bar\Pb$
production we employ explicit Feynman-diagrammatic representations of
the one-loop amplitudes and numerical reduction of tensor
integrals~\cite{Denner:2002ii}.  The factorisation of
colour matrices, the algebraic reduction of helicity structures, and
the systematic recycling of a multitude of common
subexpressions---both inside individual diagrams and in tensor
integrals of different diagrams that share common
sub-topologies---strongly mitigate the factorial complexity that is
inherent in Feynman diagrams and lead to a remarkably high CPU
efficiency.  
%
Our results have been confirmed with the 
{\sc HELAC-1LOOP} implementation of the OPP
method~\cite{Ossola:2006us,vanHameren:2009dr,Czakon:2009ss}
within the statistical Monte Carlo error of 0.2\%~\cite{Bevilacqua:2009zn}.

\section{Outline of the calculation}

In LO, the hadronic production of $\Pt\bar\Pt\Pb\bar\Pb$ proceeds via
the partonic processes $q\bar q\to\Pt\bar\Pt\Pb\bar\Pb$ and
$\Pg\Pg\to\Pt\bar\Pt\Pb\bar\Pb$, which are described by 7 and 36 tree
diagrams, respectively.  The corresponding virtual NLO QCD corrections
involve 188 and 1003 one-loop diagrams.  The real emission
contributions comprise the crossing-symmetric channels $q\bar
q\to\Pt\bar\Pt\Pb\bar\Pb \Pg$, $q\Pg\to\Pt\bar\Pt\Pb\bar\Pb q$, and
$\Pg\bar q\to\Pt\bar\Pt\Pb\bar\Pb \bar q$, which involve 64 tree
diagrams each, and the channel $\Pg\Pg\to\Pt\bar\Pt\Pb\bar\Pb \Pg$
with 341 diagrams.  Each of these contributions has been worked out
twice and independently, resulting in two completely independent
computer codes.

The virtual corrections are calculated in the Feynman-diagrammatic
approach. The diagrams are generated with two independent
versions of {\sc FeynArts} \cite{Kublbeck:1990xc,Hahn:2000kx} and
algebraically simplified with two in-house {\sc Mathematica} programs
that generate {\sc Fortran77} code in a fully automatised way.  One of
the two programs relies on {\sc FormCalc}~5.2~\cite{Hahn:1998yk} for
preliminary algebraic manipulations. The virtual corrections are
obtained from the interference of the one-loop and LO matrix elements
\looseness -1
on a diagram-by-diagram basis.

Owing to colour factorisation for individual (sub)diagrams colour sums
can be performed very efficiently. The colour-summed result is given
by a combination of previously computed colour--Born interference
terms.  This requires {\em a single evaluation} of the non-trivial
colour-stripped amplitude of each (sub)diagram.

Helicity structures are handled in a similar way.  The
helicity-dependent parts of all diagrams are reduced to a common basis
of so-called Standard Matrix Elements (SMEs), and helicity sums are
performed once and for all at the level of the SMEs--Born
interference.  The diagram-independent treatment of the
helicity-dependent parts of loop graphs is made possible by the
covariant decomposition of tensor integrals.

The one-loop amplitudes are expressed as linear combinations of
tensor-integral coefficients.  The latter are
evaluated by two independent {\em numerical\/} {\sc Fortran} libraries
that recursively reduce them to master integrals using the methods of
\citere{Denner:2002ii}.  Avoiding an explicit reduction
of analytic expressions to master integrals, this numerical approach
prevents prohibitively large expressions and permits to adapt the
reduction strategy to the specific numerical problems that appear in
different phase-space regions. An automatic cache system is
implemented that strongly boosts the reduction by recycling a
multitude of tensor integrals among Feynman diagrams with common
sub-topologies.

Ultraviolet (UV) divergences are regularized dimensionally throughout,
but infrared (IR) divergences are treated in different variants, which
comprise pure dimensional regularization with strictly massless light
quarks and a hybrid scheme with small quark masses.  The corresponding
scalar integrals are evaluated using the methods and results of
\citere{'tHooft:1979xw,Denner:2010tr}, and different
regularization schemes are translated into each other as described in
\citere{Dittmaier:2003bc}.

The treatment of rational parts is greatly simplified by the fact that
rational terms resulting from $1/\epsilon$ and $1/\epsilon^2$ poles of
IR kind vanish in truncated one-loop amplitudes
\cite{Bredenstein:2008zb}.  Rational terms arising from
UV poles of tensor integrals with $D$-dependent coefficients are
automatically extracted by means of a catalogue of residues.

\newcommand{\QG}[5]{
\gamma^{\mu_{#1}}
\gamma^{\mu_{#2}}
\gamma^{\mu_{#3}}
\gamma^{\mu_{#4}}
\gamma^{\mu_{#5}}
}
\newcommand{\TG}[3]{
\gamma^{\mu_{#1}}
\gamma^{\mu_{#2}}
\gamma^{\mu_{#3}}
}
\newcommand{\SG}[1]{
\gamma^{\mu_{#1}}
}
\newcommand{\GTE}[2]{
  g^{\mu_{#1}\mu_{#2}} } 
The reduction to SMEs is performed in such a way that no spurious
poles are generated that might cause numerical instabilities.  It
starts with process-independent $D$-dimensional relations such as
momentum conservation, Dirac algebra, transversality, and gauge-fixing
conditions for the gluon-polarisation vectors.  Once rational terms
are extracted, we further reduce SMEs
with two alternative algorithms in four dimensions.
 For the gluon induced process, the first algorithm splits each fermion chain 
into two contributions,
$
u(p_{i})
=
\sum_{\lambda=\pm}\omega_\lambda
u(p_{i}),
$  via insertion of chiral projectors 
$\omega_\pm=(1\pm\gamma^5)/2$.
This permits to employ various relations of 
type
$
{
\gamma^\mu}
{\gamma^\alpha
\gamma^\beta}
\omega_\pm
\otimes
{
\gamma_{\mu}}
=
{
\gamma^\mu}
\omega_\pm
\otimes
\left(
\gamma_{\mu}
\gamma^\beta
\gamma^\alpha
\omega_\pm
+
\gamma^\alpha
\gamma^\beta
\gamma_{\mu}
\omega_\mp
\right)$,
which 
connect Dirac matrices of 
different fermion chains~\cite{Bredenstein:2008zb,Denner:2005fg},
to reduce the full amplitude to 502 SMEs~\cite{Bredenstein:2010rs}.
Besides this procedure, which depends on process-specific aspects, we
implemented a simple process-independent reduction based on a single
four-dimensional identity of type
$\QG{1}{2}{3}{4}{5}=\GTE{1}{2}\TG{3}{4}{5}-\GTE{1}{2}\GTE{3}{4}\SG{5}
+ \mbox{perm.}$, which eliminates spinor chains with more than three
Dirac matrices without introducing
$\gamma_5$~\cite{Bredenstein:2010rs}.  This leads to 970 SMEs.
In spite of the factor-two difference in the number of SMEs, the
numerical codes based on the two different reductions have the
same---and remarkably high---CPU speed: about $180\,$ms per phase-space
point.  Thus, the obtained CPU performance, at least for this process,
does not depend on process-dependent optimisations.

To handle singularities in the real corrections we employed the dipole
subtraction method~\cite{Catani:2002hc}, in particular the {\sc
  MadDipole} implementation \cite{Frederix:2008hu} in one of our
calculations.  The $2\to 5$ matrix elements were generated with {\sc
  Madgraph}~\cite{Alwall:2007st} and checked against analytic
calculations with the Weyl--van der Waerden spinor formalism and
in-house code based on off-shell recursions.
More details are given in~\citere{Bredenstein:2010rs}.

\section{Predictions for the LHC}
\label{se:numres}

We study the process $\Pp\Pp\to\Pt\bar\Pt\Pb\bar\Pb+X$ at
$\sqrt{s}=14\TeV$ with $\Mt=172.6\GeV$ and mass\-less $\Pb$~quarks.
Massless final-state partons with rapidity--azimuthal-angle separation
$\sqrt{\Delta\phi^2+\Delta y^2}<D=0.4$ are recombined into jets using
a \mbox{$k_{\rT}$-algorithm}, and we require two b jets with
$p_{\rT,\Pb}>20\GeV$ and $|y_\Pb|<2.5$.  We use the CTEQ6 set of PDFs
but neglect the suppressed contributions from b~quarks in the initial
state.  More details are given in~\citere{Bredenstein:2010rs}

In all recent ATLAS studies of $\Pt\bar\Pt\PH (\PH\to\Pb\bar\Pb)$
\cite{Aad:2009wy,Kersevan:2002vu,Cammin:2003} the signal and its
$\Pt\bar\Pt\Pb\bar\Pb$ background were simulated by setting the
renormalisation and factorisation scales equal to half the threshold
energy, $E_{\mathrm{thr}}=2 \Mt+m_{\Pb\bar\Pb}$.  
In~\citere{Bredenstein:2009aj} we found that for this scale choice
the NLO corrections to
$\Pp\Pp\to\Pt\bar\Pt\Pb\bar\Pb$ are close to a factor 
of two.  This enhancement is due to the fact that
$\Pp\Pp\to\Pt\bar\Pt\Pb\bar\Pb$ is a multi-scale process involving
various scales well below $E_{\mathrm{thr}}/2$.
The inspection of differential distributions reveals that the cross
section is saturated by b quarks with $p_{\rT,\Pb}\ll \Mt$.  
Therefore
we introduced in \citere{Bredenstein:2010rs} the dynamical scale
\beq\label{centralscale}
\mu^2_0=\Mt\sqrt{p_{\rT,\Pb}p_{\rT,\bar\Pb}},
\eeq
which improves the perturbative convergence and minimises NLO effects
in the shape of distributions.

Using the scale \refeq{centralscale}, we discussed in
\citere{Bredenstein:2010rs} the kinematic region
$m_{\Pb\bar\Pb}>100\GeV$ and found that for all
distributions considered the NLO corrections are at the level of
20--30\% and have relatively little impact on the shape of
distributions.  On the other hand we the corrections still induce significant
distortions of the kinematic distributions in the regime of a highly
boosted Higgs boson, which offers better perspectives to observe the
$\Pt\Ptbar\PH$ signal.  Here we provide some distributions in this
scenario with $p_{\rT,\Pb\bar\Pb}>200\GeV$.

\begin{figure*}[tb]
\hfill
\includegraphics[bb= 80 455 295 655, width=.40\textwidth]       
{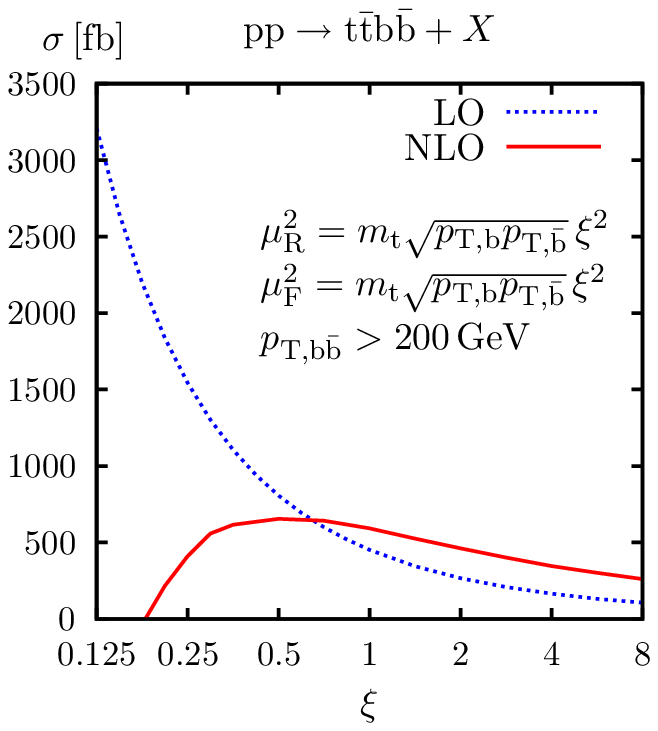}
\hfill
\hfill
\includegraphics[bb= 80 455 295 655, width=.40\textwidth]
{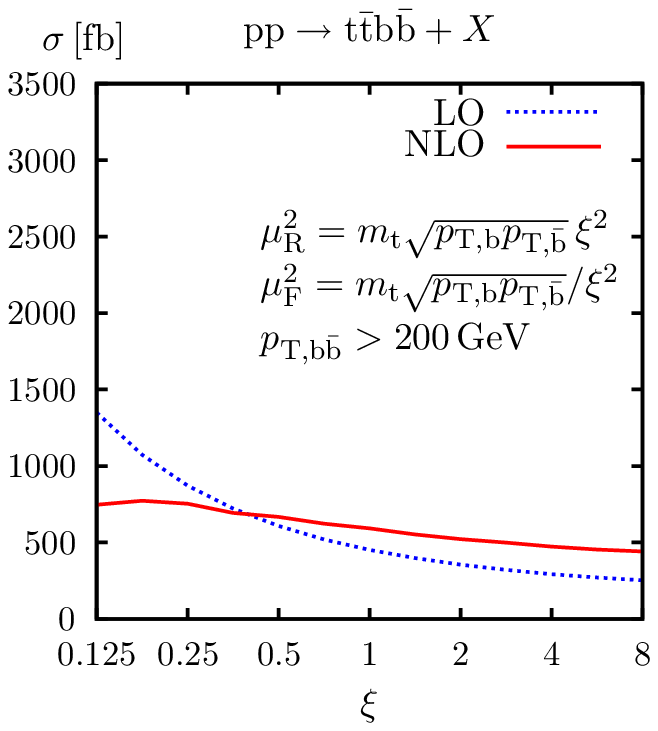}
\hfill
\vspace*{-1.30em}
\caption{Scale dependence of the LO and NLO 
  $\Pp\Pp\to\Pt\bar\Pt\Pb\bar\Pb+X$ cross section. 
 The left and the right plots describe uniform
  ($\xi_{\mathrm{R}}=\xi_{\mathrm{F}}=\xi$) and antipodal
  ($\xi_{\mathrm{R}}=\xi_{\mathrm{F}}^{-1}=\xi$) scale variations,
  respectively.  }
\label{fig:sigma_tot_2}
\end{figure*}

In \reffi{fig:sigma_tot_2} we show the scale dependence of the LO and
NLO integrated cross sections. 
Renormalisation
($\mu_{\mathrm{R}}$) and factorisation ($\mu_{\mathrm{F}}$) scales are
varied around
the central value \refeq{centralscale},
\beq       
\mu_{\mathrm{R}}=\xi_{\mathrm{R}}\mu_0,\qquad
\mu_{\mathrm{F}}=\xi_{\mathrm{F}}\mu_0.
\eeq
in a uniform ($\xi_{\mathrm{F}}=\xi_{\mathrm{R}}$) and antipodal
($\xi_{\mathrm{F}}=\xi_{\mathrm{R}}^{-1}$) way in the range $1/8\le
\xi_{\mathrm{F}},\xi_{\mathrm{R}} \le 8$.  At the central scale we
obtain $\sigma_{\mathrm{LO}}=451.8(2)\fb$ and
$\sigma_{\mathrm{NLO}}=592(4)\fb$ corresponding to
$K=1.31$.   The shape of the scale-dependence curves
indicates good convergence and stability of the perturbative
expansion.  The shifts induced by factor-two variations of the QCD
scales amount to 79\% in LO and 22\% in NLO.

For distributions we provide LO and NLO predictions with uncertainty
bands for factor-two uniform scale variations, which have a larger
impact as antipodal variations. More precisely, all observables are
evaluated at three different scales:
$\xi_{\mathrm{F}}=\xi_{\mathrm{R}}=0.5,1,2$.

\begin{figure*}[tb]
\hfill
\includegraphics[bb= 80 455 295 655, width=.40\textwidth]
{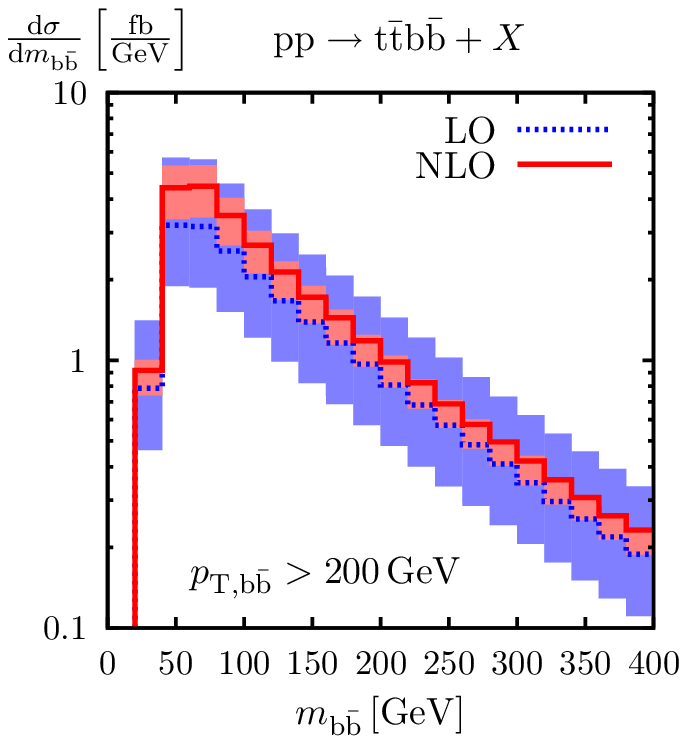}
\hfill
\hfill
\includegraphics[bb= 80 455 295 655, width=.40\textwidth]
{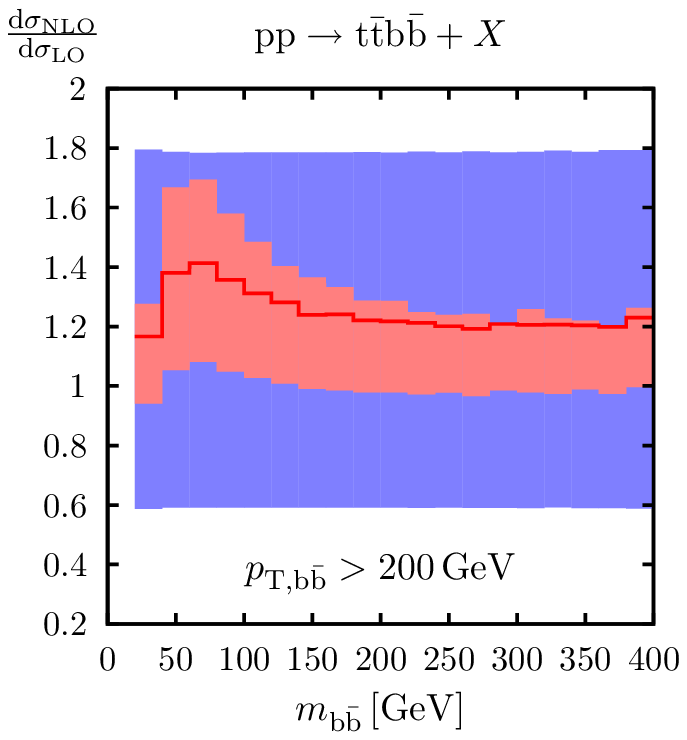}
\hfill
\vspace*{-1.30em}
\caption{Invariant-mass distribution of the
$\Pb\bar\Pb$ pair: absolute
LO and NLO predictions (left) and NLO  $K$ factor (right).
}
\label{fig:mbb_dist_2}
\end{figure*}
The $\Pb\bar\Pb$ invariant-mass distribution is displayed in
\reffi{fig:mbb_dist_2}. 
The NLO corrections induce an
appreciable shape distortion of about 20\%, in particular near the
physically interesting region of $m_{\Pb\bar\Pb}\sim 100\GeV$.  Such
an effect tends to mimic a Higgs signal and should be carefully taken
into account in the $\Pt\bar\Pt\PH(\PH\to\Pb\bar\Pb)$ analysis.

\begin{figure*}[tb]
\hfill
\includegraphics[bb= 80 455 295 655, width=.40\textwidth]
{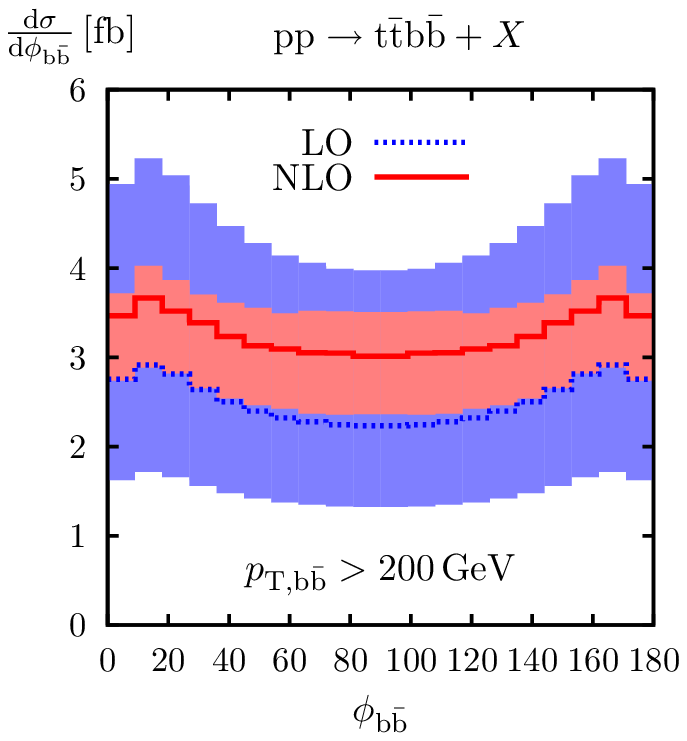}
\hfill
\hfill
\includegraphics[bb= 80 455 295 655, width=.40\textwidth]
{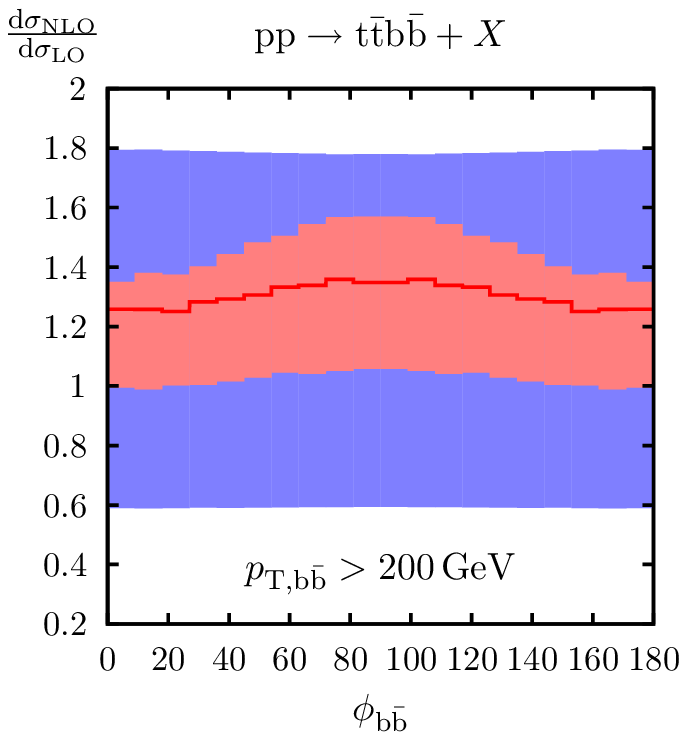}
\hfill
\vspace*{-1.30em}
\caption{Azimuthal orientation of the 
  b jets in the plane perpendicular to the $\Pb\bar\Pb$ system:
 absolute LO and NLO predictions
  (left) and NLO $K$ factor (right).  
}
\label{fig:ptbb_dist_2}
\end{figure*}

\begin{figure*}[tb]
\hfill
\includegraphics[bb= 80 455 295 655, width=.40\textwidth]
{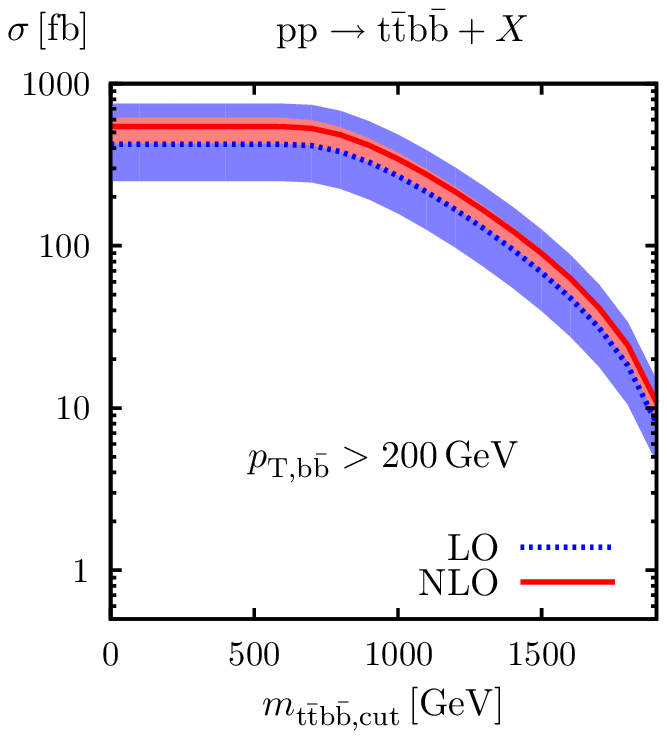}
\hfill\hfill
\includegraphics[bb= 80 455 295 655, width=.40\textwidth]
{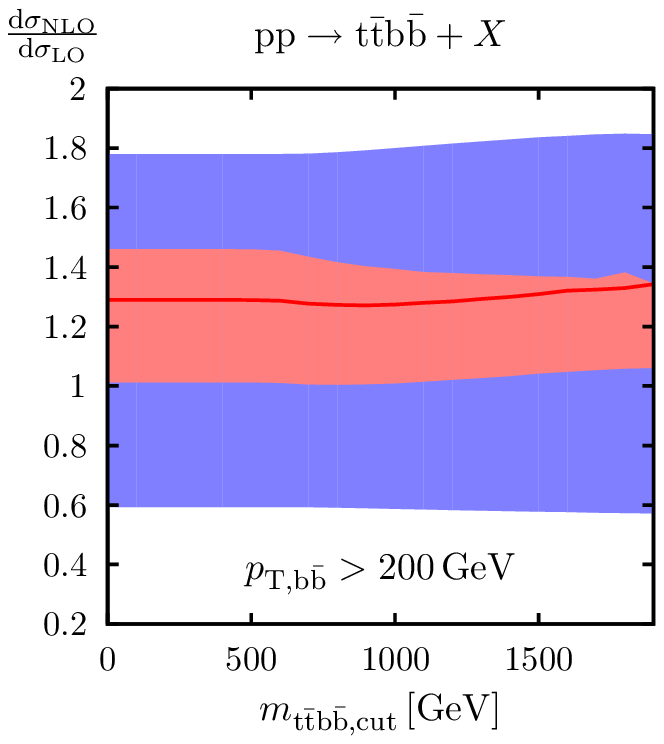}
\hfill
\vspace*{-1.30em}
\caption{Dependence of the
  cross section with respect to a cut on the $\Pt\bar\Pt\Pb\bar\Pb$
  invariant mass
  ($m_{\Pt\bar\Pt\Pb\bar\Pb}>m_{\Pt\bar\Pt\Pb\bar\Pb,\mathrm{cut}}$):
absolute LO and NLO predictions
  (left) and NLO $K$ factor (right).
}
\label{fig:mttbbcut_dist_2}
\end{figure*}

For other distributions the shape distortion is not as sizeable. 
As examples we show the distribution in the azimuthal angle
$\phi_{\Pb\bar\Pb}$ that  represents the azimuthal orientation
of the b jets with respect to the beam direction in the plane
perpendicular to the $\Pb\bar\Pb$ momentum in \reffi{fig:ptbb_dist_2}
and the dependence of the
  cross section with respect to a cut on the $\Pt\bar\Pt\Pb\bar\Pb$
  invariant mass in \reffi{fig:mttbbcut_dist_2}.


\section{Conclusion}
The observation of the $\mathrm{t\bar t H(H\to b\bar b)}$ signal at
the LHC requires a very precise description of the $\mathrm{t\bar t
  b\bar b}$ irreducible background. The NLO QCD corrections reveal
that the scale choice adopted in previous LO simulations of
$\mathrm{pp\to t\bar t b\bar b}$ does not account for the multi-scale
character of this process and underestimates its cross section by a
factor of two. A suitably chosen dynamical scale significantly reduces
both the $K$ factor and the residual NLO scale uncertainty.  For
standard 
cuts NLO effects feature a relatively small
kinematic dependence, but in the regime of highly boosted Higgs bosons
significant distortions are still present in the shape of some
distributions.

\end{document}